\documentstyle[preprint,aps,epsf,pre]{revtex}
\begin{document}
\draft

\title{Some stochastic phenomena in a driven double-well system}

\author{Mangal C. Mahato\footnote{Present address, Department of Physics, Guru Ghasidas University,
Bilaspur-495009 (M.P.), India.}
 and A. M. Jayannavar\footnote{Corresponding author, e-mail: jayan@iopb.ernet.in} \\
Institute of Physics, Sachivalaya Marg,
Bhubaneswar-751005, India\\}

\maketitle

\begin{abstract}
We study the overdamped motion of a Brownian particle in a
driven double-well system to understand various physical
phenomena observed experimentally. These phenomena include
hysteresis, stochastic resonance, and net unidirectional motion
in a symmetric periodic system. We argue that the area of the
hysteresis loop so defined is a good measure of synchronization
(with respect to the applied field) of passages between the two
wells. We find that stochastic resonance may be relevant even in
case of large amplitude driving field due to a recently
discovered phenomena of noise induced stability of unstable
states. We try to find relation between some of these apparently
different phenomena.
\end{abstract}

\section{Introduction}

There are several phenomena in physics and chemistry which are
simply seen as an event of potential barrier crossing. The
examples include the motion of point defects in crystalline
solids, transitions in bistable optical devices, and chemical
reactions. The event of crossing over the potential barrier is
aided by fluctuating forces and the problem was formally dealt
with initially by Kramers\cite{Kramers}. The transition from one
local equilibrium situation to the other finds a simple
theoretical description in the study of two-well potential
systems. In these nonlinear deterministically bistable systems
the inclusion of noise gives rise to interesting dynamical
effects. The stochastic dynamics in a static two-well potential
has been studied extensively and is fairly well
understood\cite{RMP}. However, it is very difficult to find
analytical solutions when the system is driven by an external
field. We study the driven two-well system numerically as a
simple model to understand various experimentally observed
phenomena such as hysteresis, stochastic resonance\cite{Fauve},
and net unidirectional motion in a symmetric periodic
potential\cite{Svoboda}.

When the response of a system lags behind an externally applied
field the system is said to show hysteretic behaviour {\it
vis-a-vis} the applied field. The most familiar example of
hysteresis being the variation of magnetization of a
ferromagnetic system as an external magnetic field is varied in
time. The phenomenon, however, is quite general and any system
that shows metastability exhibits hysteresis. In the present
work we treat it as a result of competition between various
rates\cite{Gilmore} (time scales).  In particular, in the
two-well system we have the rate of relaxation at the bottom of
the wells, the mean rate at which passage takes place between
the two wells across the potential barrier and the rate at which
the system is being driven by the external field; the former two
rates depending on the external field. Depending on the
situation the external field could be taken as varying
monotonically in time or varying periodically.  In order to
assign a single rate (or at most two rates) of variation of the
external field we either consider a linear sweep\cite{JSP}
(monotonic) or a saw-tooth type of (periodic) field
sweep\cite{SR}. In the case of periodic field sweep we show
(section II) that hysteresis loop area shows maximum as a
function of the strength of the fluctuating forces. Moreover, we
argue that the hysteresis loop area is a good measure of degree
of synchronization of passages from one well to the other. Thus,
the more the passages get synchronized with the applied field
the larger is the hysteresis loop area. The same numerical data,
namely the distribution of field values (or time intervals) at
which passages take place from one well to the other, are used
to understand the rest of the phenomena listed in the previous
paragraph.

In the recent past stochastic resonance has been studied
extensively. Initially it was discovered theoretically to
understand the reccurrence of ice ages\cite{Benzi} at certain
regular intervals of time ($\approx 10^5$ years). Subsequently
the phenomenon was observed experimentally\cite{Fauve} in
various systems such as electronic trigger circuits, two-mode
ring lasers, and mammalian neuronal networks.  The optimization
of the response of a system to an input signal as a function of
the input noise strength is termed as stochastic resonance. It
is a nonlinear effect. The response of the system is measured as
the ratio of output signal (at the input signal frequency) to
(the output background) noise strength. This signal-to-noise
ratio (SNR) (obtained from the power spectrum of the output
signal) shows a peak as a function of input noise strength. For
a simple understanding the nonlinear system can be modeled as a
two-well potential. The residence time distribution of a
Brownian particle in one of the two wells, as the system is
driven by an input periodic signal, is then taken as the
response. The simple model system shows stochastic resonant
behaviour (section III) given a small amplitude periodic signal.
This resonance is seen to occur as a result of synchronized
passages across the potential barrier in the low "frequency"
limit establishing a close relationship between these two
phenomena.  Stochastic resonance, however, can also be seen in
case of large amplitude (overcritical) input periodic signals.

The two-well system can be represented by the usual Landau
potential (Fig. 1). When the strength of the external field
$h(t)$ becomes larger (Eqs. 2 and 3) than a critical value
$h_c$ one of the two wells disappears and one single well
remains. Therefore a Brownian particle with its instantaneous
position at the now nonexistent well will roll down towards the
single remaining well. When the large amplitude field is
periodic each well in turn will disappear and then reappear
exactly once in a period. Therefore, if the field sweep is slow
enough a Brownian particle should pass from one well to the
other exactly once in a period when the noise strength is almost
zero. The scenario, however, changes as the strength of the
noise is gradually increased. Instead of the average number of
passages per cycle (ANPPC) increasing from one it, in fact,
decreases initially and then on increasing the noise strength
further it attains a minimum value and then begins to increase
and ultimately becoming larger than one corresponding to the
case of noise dominated dynamics. Similar effects have been
predicted recently in other systems also\cite{Dayan}. This
decrease of the ANPPC from 1 as the noise strength is increased
from zero (or the noise induced stability of an unstable state)
helps to obtain stochastic resonance in the case of large
amplitude input signals (section IV).

The phenomena discussed in the last two paragraphs were studied
taking the external periodic field sweep to be symmetric in
time.  The saw-tooth type periodic field sweep can easily be
taken as temporally asymmetric keeping the average force per
cycle, due to the external field, zero. We propose a physical
model for a net preferrentially unidirectional motion of a
Brownian particle in a symmetric periodic potential when the
system is subjected to such a temporally asymmetric periodic
field. The fluctuating forces (noise) necessary to achieve such
a motion need not be colored.  A symmetric Gaussian white noise
will be sufficient. We provide numerical support to this model
by studying the two-well system and calculating the net
accumulation of particles in each of the two wells when
subjected to temporally asymmetric saw-tooth type periodic
fields (section V). We further show that such a physical model
can be made more efficient by suitably choosing the strength of
the fluctuating forces. The motivation for the proposition of
such a model comes from an experimentally observed phenomenon of
net (statistically) preferrentially unidirectional motion of
protein motor (kinesin) molecules along
microtubules\cite{Svoboda}. And the idea\cite{IJP} of using two
time scales for the field sweep comes from the two flagellar
strokes (power and reverse) executed by sperm cells for their
propulsion. This subject is being persued intensively in recent
years under the title of correlation ratchets, thermal ratchets
or the Maxwell's demon-type information engines.

\section{Hysteresis}

We calculate hysteresis from the first-passage-time (FPT)
distribution of a Brownian particle in the double-well potential
under the influence of an external field. We also calculate it
from the residence time distributions in each of the two wells.
The motion of the particle is studied by solving, numerically,
the overdamped Langevin equation
\begin{equation}
\dot m=-\frac{\partial \Phi}{\partial m}+\hat f(t),
\end{equation}
where m is the order parameter (in this case position of the
particle) and the dot over it denotes derivative with respect to
time.
\begin{equation}
\Phi (m,t)=U(m)-mh(t),
\end{equation}
(Fig. 1) where, $U(m)$ is the Landau two-well symmetric
potential
\begin{equation}
U(m)=-\frac{a}{2}m^2+\frac{b}{4}m^4
\end{equation}
and $h(t)$ is the external time dependent field sweep. 
$\hat f(t)$ are randomly fluctuating forces and are taken
to be Gaussian with statistics
\begin{equation}
<\hat f(t)>=0,
\end{equation}
and
\begin{equation}
<\hat f(t)\hat f(t')>=2D\delta (t-t').
\end{equation}
Here $<...>$ represents an ensemble average over all
realizations of the random forces.

\subsection{The first-passage times}

In the beginning (at time $t=0$) we take the Brownian particle
at the bottom of one of the two wells and then monitor its
motion as time progresses. As soon as the particle reaches the
bottom of the other well we record the time $t=\tau$ and stop
the process. We begin the process all over again from the bottom
of the first well and repeat the earlier procedure. By repeating
the process for a large number of times (for meaningful
averages) we obtain the density distribution $\rho (\tau)$ of
the FPT's, $\tau$. $\rho (\tau)$ depends on how the field $h(t)$
is swept and the strength of the fluctuating forces $D$.

When the field $h(t)$ is swept linearly,
\begin{equation}
h(t)=h_0+\dot ht,
\end{equation}
where $\dot h$ is a fixed constant. By taking, for example, $h_0
\gg h_c$ (the critical field $=2(a/3)^{3/2}$ for $b=1$; we take
$a=2$ in our numerical calculation throughout) and $\dot h<0$
and start from the bottom of the right side well one can obtain
$\rho (\tau)$ and get a sensible hysteresis loop $M(h)$.
\begin{equation}
\frac{M(h)}{h_c}=1-2 \int_{h}^{h_0} \rho (h') dh',
\end{equation}
where $\rho (h)$ is the distribution of field values at
$\tau$'s. Equation (7) gives the upper half of the hysteresis
loop. The other half is obtained by symmetry\cite{JSP}. The
linear sweep of the field has physical analogues in, for
example, sweeping the temperature $T$ (as a function of time)
from above the freezing temperature of a melt to down below that
temperature, where $M(T)$ would then correspond to the degree of
nucleation of crystallites in the melt. However, $D$ would now
depend on $T$.  By suitably modelling $D(T)$ one can obtain the
qualitative features of the usual (T-T-T) curves\cite{Uhlman}.
Also it should be noted that by taking $\dot h$ small one can
check whether the hysteresis loop area so obtained follows any
power-law behaviour\cite{Rao} in terms of $\dot h$. In an
earlier work we, however, do not find any universal power law
behaviour\cite{JSP}. 

When the field $h(t)$ is varied periodically (in our case as a
saw-tooth) in time the distribution $\rho (\tau)$ shows (Fig. 2)
gradually diminishing peaks occurring periodically\cite{SR}. The
periodicity roughly matches the periodicity of $h(t)$. Since the
field oscillates between $h_0$ and $-h_0$, we can again find the
distribution $\rho (h)$ of passage fields $h(\tau)$ and the
hysteresis loop $M(h)$. We have shown earlier that the
hysteresis loop area acquires a maximum as a function of $\dot
h$ as well as $D$\cite{SR}. The area of the hysteresis loop as
defined in the present work (see below) is a good measure of the
degree of synchronization of passages (from one well to the
other) with respect to the external field sweep $h(t)$.

For periodic $h(t)$ with amplitude $h_0<\vert h_c\vert $, a
Brownian particle on the right well (Eq. 2) will encounter the
least barrier of passage to the left well when $h=-h_0$.
Therefore, the probability of passage from the right well to the
left well will be the largest when $h(t)=-h_0$. If we consider
passage to take place only at $h(t)=-h_0$ from the right well to
the left and only at $h(t)=h_0$ from the left well to the right,
we have the case of perfect synchronization of passages with
respect to the variation of the field. This will, however,
correspond to $\rho (\tau)$ having sharp peaks occurring
periodically at $t: h(t)=-h_0$ if we begin from the right well.
And the corresponding $\rho (h)$ will have a single sharp peak
at $h=-h_0$. Therefore, from Eq. 7 we will have a rectangular
hysteresis loop and hence with the largest possible loop area.
On the other hand if the passages are completely random (the
case of least synchronization) we will have a uniform $\rho (h)$
from $h_0$ to $-h_0$ and the hysteresis loop will consequently
have the smallest area (zero). From the discussions of these two
extreme cases we arrive at the conclusion that hysteresis loop
area, indeed, provides a good measure of degree of
synchronization of passages from one well to the other as the
field is periodically swept. Therefore the earlier
result\cite{SR} shows that the passages become most synchronized
at optimum values of noise strength $D$ and field sweep rate
$\dot h$.  Similar results are obtained when hysteresis is
calculated from the residence time distribution (see below).

\subsection{Residence time distributions}

Instead of studying a large number of identical replicas of a
system one can study the same system for a long time and observe
the passages a Brownian particle executes in the course of time.
This way one can find the residence times and their
distributions $\rho_1(\tau)$ and $\rho_2(\tau)$, in each of the
two (left and right) wells, respectively, (Fig. 3). Also one can
find the passage field distributions $\rho_{12}(h)$ and
$\rho_{21}(h)$ for passages from the left well to the right well
and from right well to the left well, respectively. However, the
first-passage field distribution $\rho (h)$ from right well to
the left well, for example, is not expected to be exactly the
same as $\rho_{21}(h)$. This is because for FPT calculation one
starts from the same initial condition $h(t=0)$, whereas for
$\rho_{21}(h)$ $\rho_{12}(h)$ acts as the distribution of
initial field values, $\it etc$. Also, in the previous (FPT
based) work $\rho (h)$ was obtained for variation from $h_0$ to
$-h_0$. However, the proper book-keeping of the field values of
passages from one well to the other throughout the cycle of
$h(t)$, namely from $h_0$ to $-h_0$ and then back to $h_0$, is
required. Taking all these aspects into account we calculate the
probability of the Brownian particle being in the right well
($m_2(h)$) when the field value is h:
\begin{equation}
m_2(h)=m_2(h-\Delta h)-m_2(h-\Delta h) \rho_{21}(h)\Delta h
+m_1(h-\Delta h)\rho_{12}(h)\Delta h,
\end{equation}
and similarly for $m_1(h)$. The difference $m(h)=m_2(h)-m_1(h)$
is analogous to magnetisation in magnetic systems. The
stationary (closed) hysteresis loop $m(h)$ is obtained by
iteration. Note that there is a qualitative difference in the
hysteresis loops calculated using the two methods. In the
earlier case (FPT based) the hysteresis loop was always
saturated by construction. In the present case the hysteresis
loop need not be saturated and indeed one obtains saturation
only for $h_0>\vert h_c\vert$ together with small $\dot h$ and
large $D$ cases (Fig. 4). The area of hysteresis loop so
obtained shows a maximum\cite{Relation} as a function of $D$ and
also as a function of $\dot h$ (Fig. 5). As in case of
hysteresis loop area from first-passage field distribution (Eq.
7) one can argue that in this case too the hysteresis loop area
is a good measure of degree of synchronization of passages. Thus
the passages, again, become most synchronized at an optimum
value of the noise strength $D$ and also at an optimum value of
$\dot h$.

\section{Stochastic resonance}

We use the residence time distributions $\rho_1(\tau)$ or
$\rho_2(\tau)$ to study stochastic resonance. As mentioned in
the introduction the signal-to-noise ratio (SNR) in the power
spectrum of these distributions is used to study it. We adopt
the following procedure to obtain the power spectrum\cite{NR}.
We calculate the histogram of ,say, $\rho_2(\tau)$ (Since the
periodic drive is symmetric the two distributions are
identical.) dividing the period $T$ of $h(t)$ into 10 equal
parts. We assign the mid-point of each time interval
$\Delta=T/10$ the value equal to the height of the histogram at
the interval. Usually the number of such data points is not
large. We augment the data points obtained from the histogram by
zeroes at similar equal intervals so that the maximum number of
paded data points, unless mentioned otherwise, is 16384
($=2^{14}$). We then calculate the Fourier transform of this
data using the Fast-Fourier-Transform (FFT) routines. The number
of (doctored) data points is taken to be large in order to
obtain Fourier trasforms at close intervals of frequency. Square
the suitably normalized Fourier transformed data to give the
power spectrum of $\rho (\tau)$.  The normalization is done
correctly such that the spectral density is one for zero
frequency. We obtain sharp peaks at frequencies equal to
$\frac{2\pi n}{T/2}$ for $n=1,2,\ldots$. We measure the first
peak height and the background value of the spectrum at the same
frequency. The background value of the spectrum is usually small
and the possibility of commiting error in assigning a value to
it is correspondingly large. The background value of spectrum is
termed as noise and the peak height as the signal. We thus
calculate the SNR for each value of ($\dot h, D$). The SNR shows
maximum as a function of $D$ as well as $\dot h$ (Fig. 6). It
should be noted that the measurement of SNR is not very accurate
(large errors) but their variation as a function of $D$ or $\dot
h$ is so large that the trend of their variation as a function
of $D$ or $\dot h$ is not affected.  (Note that our conclusions
pertain mostly to the trends of variation rather than to the
actual values.)

The maximum of SNR as a function of $D$ indicates stochastic
resonance. We find the value of $D=D^m_{SNR}$ at which the
maximum occurs for various values of ($h_0,\dot h$). Similarly,
we note the value $D=D^m_a$ at which the hysteresis loop area
becomes maximum for the same set of values ($h_0, \dot h$). The
plots of $D^m_a$ as well as $D^m_{SNR}$ as a function of $\dot
h$ for fixed values of $h_0$ is shown in Fig. 7. It is
noteworthy that eventhough $D^m_a$ and $D^m_{SNR}$ differ widely
for large $\dot h$, as $\dot h$ is decreased towards zero
$D^m_a$ and $D^m_{SNR}$ become closer and closer\cite{Relation}.
As mentioned earlier hysteresis loop area is a good measure of 
degree
of synchronization of passages from one well to the other.
Therefore, at small $\dot h$ stochastic resonance occurs when
the synchronization is close to its largest value. This result
indicates that there is a close relationship between
synchronization and stochastic resonance. This aspect of 
stochastic resonance had not been explored so far.

SNR also peaks\cite{Relation} as a function of $\dot h$ as does
the hysteresis loop area. It indicates that there is a time
scale of maximum response of the system depending on the values
of $h_0$ and $D$. The occurrence of this maximum as a function
of frequency is closer to the conventional resonance phenomena.
There has been some controversy in literature about whether the
stochastic resonance is a bonafied resonance\cite{Bonafied} at
all. In Ref. [15] an effort has been made to resolve the 
controversy by measuring the strength of the first peak of the
distribution $\rho_{12}(\tau)$ and noting their variation. Here 
our treatment takes account of all the peaks of $\rho_{12}(\tau)$ 
in order to calculate the hysteresis loop area which is used to
measure the degree of synchronisation of passages. The maximum 
of SNR as a function of frequency (equivalent 
to $\dot h$ in our case) has been found\cite{Berdi} only
recently (analytically) to occur in a double (square)-well
potential system.

\section{Noise-induced stability of unstable states}

In the previous section we studied the response of the two-well
system to a small amplitude (subcritical) input signal in order
to obtain stochastic resonance. We now study the response of the
same system to a large amplitude (overcritical) input periodic
signal. In particular, we take $h_0=1.4h_c$. As can be seen from
Eq. 2 the left well disappears and gives way to a steep
potential line for $h(t)>h_c$ and only the deepened right well
remains (Fig. 1). The reverse situation results when
$h(t)<-h_c$. We find that when $D=0$ and $\dot h$ less than a
threshold value $\dot h_{th}$ (in our case $\approx 0.384h_c$
per unit time) a particle rolls down alternately to both the
wells exactly once in a period of $h(t)$. Therefore, the average
number of passages per cycle (ANPPC) equals 1. However, as $\dot
h$ is taken larger than $\dot h_{th}$ the ANPPC goes to zero
very sharply (Fig. 8).  However, the situation changes as $D$ is
gradually increased from zero.

Figure 9 shows the variation of ANPPC as a function of $D$ for
various values of $\dot h$. For any $\dot h< \dot h_{th}$
increasing D from zero initially decreases ANPPC. It attains a
minimum value and then begins increasing ultimately becoming
larger than 1 for large $D$. For all $\dot h> \dot h_{th}$ ANPPC
monotonically increases from a small value (zero) as $D$ is
increased from zero (Fig. 9). Therefore, the range $0<\dot h<
\dot h_{th}$ provides an interesting interval within which the
passage of the Brownian particle from an unstable situation is
slowed down because of the presence of fluctuating forces. We,
therefore, see the gradual emergence of more than one peaks of
$\rho_{1(2)}(\tau)$ as $D$ is increased from zero (Fig. 10)
which on further increament of $D$ get broadened. Interestingly,
on calculating the power spectrum of $\rho_2(\tau)$ we find
corresponding evolution of broad peaks (centred at the periodic
field frequency and at its harmonics) as $D$ is increased from
zero. (No peaks for $D=0$.) However, on increasing $D$ further
the peaks gradually disappear (Fig. 11). This is clearly a
signature of stochastic resonance in this case of large
amplitude forcings.  Thus, the slowing down of the passage of
the Brownian particle in a two-well system makes it meaningful
to study SR as a response to large amplitude input signal.

The role of noise in the two cases of small and large amplitude
forcings are quite the opposite. In the large amplitude case the
observance of stochastic resonance is because the presence of 
noise slows down the decay of the unstable states whereas in the
small amplitude cases the noise helps to overcome the potential
barrier for passage. In the case of large amplitude forcings the
passages take place mostly when the intervening potential 
barrier is missing. The passages are not noise induced, rather 
they are noise controlled. Correspondingly the hysteresis will
have a different nature. For analogy, the process in this 
case will be more akin to large amplitude temperature cycling
across the freezing point of a liquid than field cycling well
within the ferromagnetic regime. The calculated hysteresis loop
area does not show maximum at the same frequencies or noise 
strengths where SR appears.
 
\section{Two-well system subjected to temporally asymmetric
periodic forcing}

It is quite contrary to common sense perception to observe
asymmetric motion of a particle in a symmetric periodic
potential without any obvious external bias. Several physical
models\cite{Magnasco} were proposed to obtain asymmetric motion
in a periodic potential. In most cases the periodic potential is
taken to have ratchetlike asymmetry within a period and some
correlated fluctuating forces are applied. We propose that
asymmetric motion can be obtained even in a symmetric
(nonratchetlike) periodic potential by subjecting the particle
to a Gaussian white noise of zero mean. This is made possible by
the application of a temporally asymmetric (within a period)
periodic forcing.  We consider a saw-tooth type periodic field
$h(t)$ but having the magnitude of the two slopes different. We
consider the external field $h(t)$ to be such as to exert a zero
mean force on an otherwise free particle implying thereby that
the particle motion is not biased on the average by the external
field.

We take the amplitude $h_0$ of the external field $h(t)$ to be
small ($h_0<h_c$) and monitor the motion of a
particle\cite{Asym} on the two-well potential subjected to white
noise (fluctuating forces). We then find the mean values
($\overline{\tau}_1$ and $\overline{\tau}_2$) of residence times
in each of the two wells. We find that when $h(t)$ is asymmetric
(defined by the asymmetry parameter $\Delta
=\frac{T_1-T_0/2}{T_0/2}$, where $T_1$ is the time the field
takes to decrease from $h_0$ to $-h_0$ and $T_0$ is the period
of $h(t)$)) the two mean values $\overline{\tau}_1$ and
$\overline{\tau}_2$ (corresponding to left and right wells,
respectively) are not equal. This simply implies that the
particle spends unequal time in each of the two wells. Moreover,
$\overline{\tau}_1+\overline{\tau}_2>T_0$. If
$\overline{\tau}_1+\overline{\tau}_2$ were equal to $T_0$ we
would conclude that eventhough $\overline{\tau}_1\neq
\overline{\tau}_2$ the particle makes on the average equal
number (=1) of passages from left to right and from right to left in
a cycle of $h(t)$. However, the inequality implies that the
average number of passages from the left well to the right is
not equal to the average number of passages from the right well
to the left per cycle of $h(t)$. This observation helps us to
make the important inference that in a periodic potential system
a similar asymmetry in passages will ensue between the two
adjacent valleys resulting in a net current across the periodic
system. It is noteworthy that the passages considered here are
in a symmetric (nonratchetlike) potential unlike in the work of
Magnasco\cite{Magnasco} who considers ratchetlike asymmetric 
potential. Also
here the passages are due to periodic forcings of amplitude 
smaller than the critical amplitude $h_c$ necessitating the 
presence of noise unlike in the work of Ajdari, et al where
large amplitude forcings are used without the noise\cite{Ajdari}.

Figure 12 shows how $m=\overline{\tau}_2-\overline{\tau}_1$ vary
as a function of $\Delta$. The monotonic rise is quite evident.
Moreover, from Fig. 13 we see that given an asymmetry parameter
$\Delta$, $m$ acquires a maximum as a function of $D$. This
shows that the net asymmetric current can be optimized by
suitably choosing the noise strength $D$.  Again, the area of
the hysteresis loops becomes maximum (Fig. 14) as a function of
$D$ for a given $\Delta$. But the position of this maximum does
not coincide with the position of the maximum of $m$ as a
function of $D$. We can also see how the hysteresis loop area
vary as $\Delta$ is changed. 

\section{Conclusion}

In this work we have tried to understand the phenomena of
hysteresis, stochastic resonance, synchronization and asymmetric
passages in a model symmetric bistable system. We find that
stochastic resonance, at least at low signal frequencies, could
be a result of synchronised response of the system to the input
periodic signal. The degree of synchronization itself being
optimized by the noise strength as well as the signal
"frequency".  Also, we obtain stochastic resonance in the
response to a large amplitude periodic forcing as a result of
noise induced slowing down of a Brownian particle rolling down
an inclined potential surface. By studying the processes in the
bistable system we could also come to the conclusion that a net
unidirectional current can be obtained even in a symmetric
(nonratchetlike) periodic system subjected to Gaussian white
noise when acted upon by a temporally asymmetric but zero
averaged periodic forcing. The unidirectional current is also
optimized by the noise strength.  It is noteworthy that our
model Landau (two-well) potential system is a very simple
nonlinear system yet its study provides valuable understanding
of all the above mentioned experimentally observed phenomena.

\begin{figure}
\caption{Plots of (a) $\Phi (m)$ as a function of $m$ for
$h(t)=0.6h_c$, and (b) $m_{xtrm}$, the extrema points of $\Phi
(m)$ as a function of $h$, for $a=2.0$, and $b=1.0$.}
\end{figure}
\begin{figure}
\caption{Shows (a) the time variation of the external
field $h(t)$, (b) the first-passage-time distribution $\rho
(\tau)$, (c) the passage field distribution $\rho (h)$, and (d)
the corresponding hysteresis loop calculated from $\rho(h)$, for
$h_0=0.7h_c$, $D=0.3$, and the period of $h(t)$, $T_0=28.0$.}
\end{figure}
\begin{figure}

\caption{Shows (a) the residence time distribution $\rho
_1(\tau)$ in the well 1, and (b) the passage field distribution
$\rho _{21}(h)$ for passage from the right well 2 to the left
well 1, for $h_0=0.9h_c$, $D=0.2$, and period of $h(t)$,
$T_0=36.0$.}
\end{figure}
\begin{figure}

\caption{Hysteresis loops calculated from $\rho_i(\tau)$ and
$\rho_{ij}(h)$, $i,j=1,2$, for various values of $h_0=1.4h_c$
(outer), $0.9h_c$(middle) and $0.7h_c$ for $\dot h=0.05h_c$ and
$D=0.2$.} 
\end{figure}
\begin{figure}

\caption{Plots of hysteresis loop area $A$ (a) as a
function of $D$ for $\dot h=0.05h_c (\circ)$, $0.1h_c (\Box)$,
$0.2h_c (\diamond)$, $0.4h_c (\bigtriangleup)$, and $0.6h_c
(\bigtriangledown)$, and (b) as a function of field sweep rate
$\dot h$ for $D=0.1 (\circ)$, $0.15 (\Box)$, $0.2 (\diamond)$,
and $0.3 (\bigtriangleup)$ for $h_0=0.9h_c$. In this and in the
rest of the figures the lines joining the points are only to
guide the eye.}
\end{figure}
\begin{figure}

\caption{Plots of signal-to-noise ratio (SNR) for
$h_0=0.9h_c$ (a) as a function of $D$, for $\dot h/h_c=0.05
(\circ)$, $0.1 (\Box)$, $0.2 (\diamond)$, $0.4
(\bigtriangleup)$, $0.5 (\triangleleft)$, $0.6
(\bigtriangledown)$, and $0.72 (\triangleright)$, and (b) as a
function of $\dot h$ for various values of $D=0.1 (\circ)$,
$0.15 (\Box)$, and $0.2 (\diamond)$.}
\end{figure}
\begin{figure}

\caption{Plots the peak positions of the plots of
hysteresis loop area $A$ versus $D (\circ)$ and those of SNR
versus $D (\Box)$, for $h_0=0.7h_c$ (empty symbols, solid
joining lines) and for $h_0=0.9h_c$ (filled symbols, dashed
joining lines).} 
\end{figure}
\begin{figure}

\caption{Average number of passages per cycle (ANPPC) is
plotted as a function of $\dot h$ for $D=0.00001$. Averaging is
done over 5000 cycles. The line is drawn to guide the eye.}
\end{figure}
\begin{figure}

\caption{Shows ANPPC as a function of $D$ for various
$\dot h$ values:(a) $0.05h_c(\circ)$, (b) $0.2h_c (\Box)$, (c)
$0.28h_c (\diamond)$, (d) $0.35h_c (\triangle)$, and (e) $0.4h_c
(\triangleleft)$. For $\dot h\widetilde{>}0.392h_c$ the curves,
for example the curve for $\dot h=.4h_c$, show monotonic
behaviour starting from ANPPC=0. The lines are drawn to guide
the eye.} 
\end{figure}
\begin{figure}

\caption{Shows $\rho(\tau)$ for $h_0=1.4h_c$, $\dot h=0.028h_c$
and $D=0.001$ (bottom), $D=0.1$ (middle; shifted up by 0.25) and
$D=0.4$ (shifted up by 0.5).}
\end{figure}
\begin{figure}

\caption{Shows power spectral density (psd) for three
values of $D$: (a) $D=.001$ (dotted), (b) $D=.1$(solid line), and
(c) $D=.4$ (dash-dotted), at $\dot h=.028h_c$. We have taken, for
each curve, 256 (augmented with zeroes) data points at a time
interval of 2.0.} 
\end{figure}
\begin{figure}

\protect\caption{Shows the variation of accumulation $M$ in a
well as a function of $\Delta$ for (a) $D=0.15 (\Box)$, (b)
$D=0.2 (\diamond)$, (c) $D=0.5 (\bigtriangleup)$, and (d) $D=0.7
(\bigtriangledown)$.} 
\end{figure}
\begin{figure}

\protect\caption{Plot of $M$ versus $D$ for (a) $\Delta=0.5
(\circ)$, and (b) $\Delta=\frac{13}{14} (\Box)$.}
\end{figure}
\begin{figure}

\protect\caption{Plot of hysteresis area as a function of $D$
for (a) $\Delta=0.5 (\circ)$, and (b) $\Delta=\frac{13}{14}
(\Box)$.} 
\end{figure}
\vfill
\newpage
\begin{figure}
\protect\centerline{\epsfxsize=6in \epsfbox{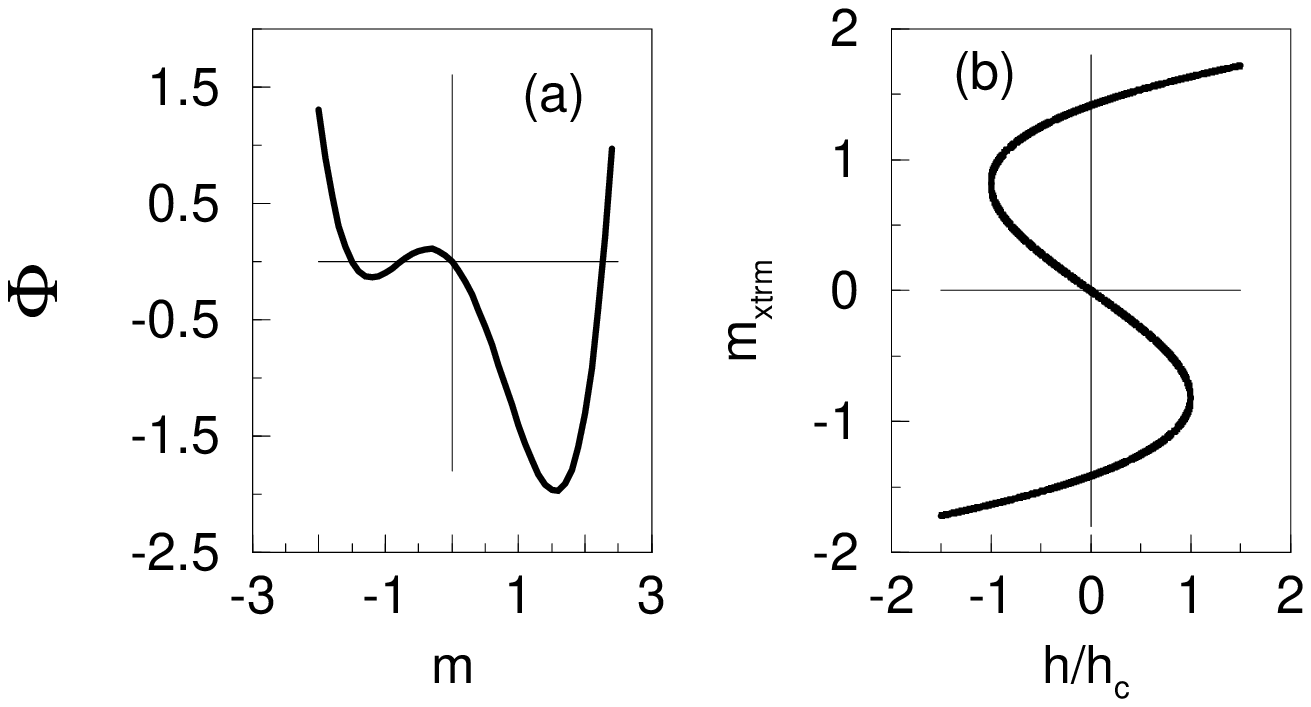}}
\end{figure}
\label{Fig.1}
\vfill
\newpage

\begin{figure}
\protect\centerline{\epsfxsize=6in \epsfbox{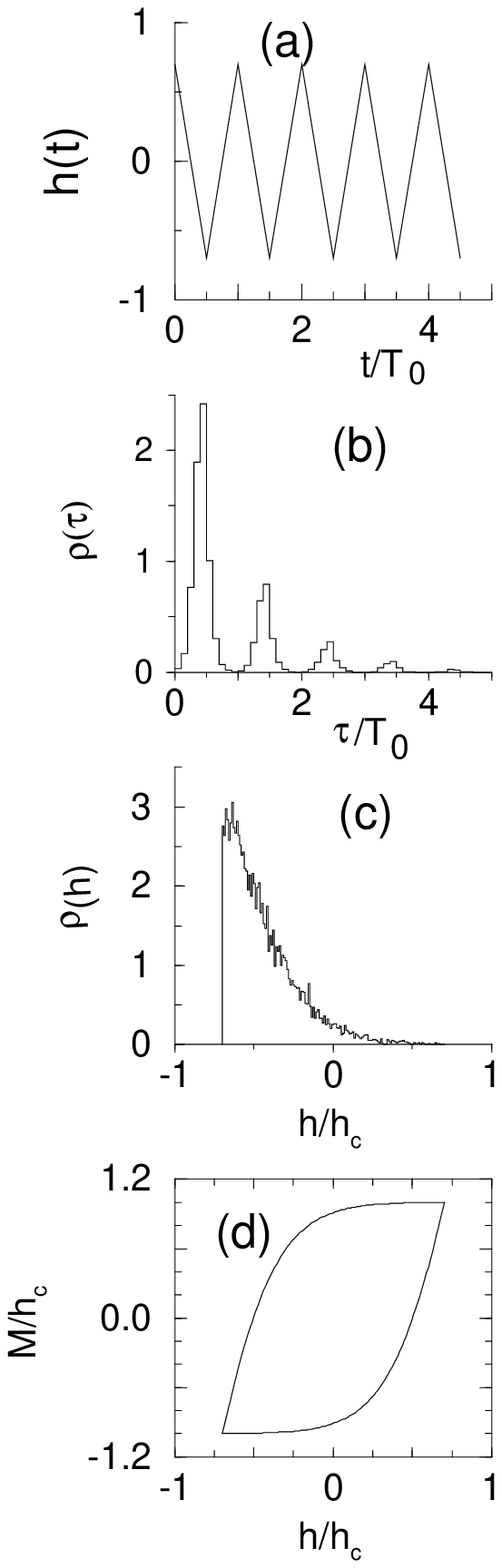}}
\label{Fig.2}
\end{figure}
\newpage
\begin{figure}
\protect\centerline{\epsfxsize=6in \epsfbox{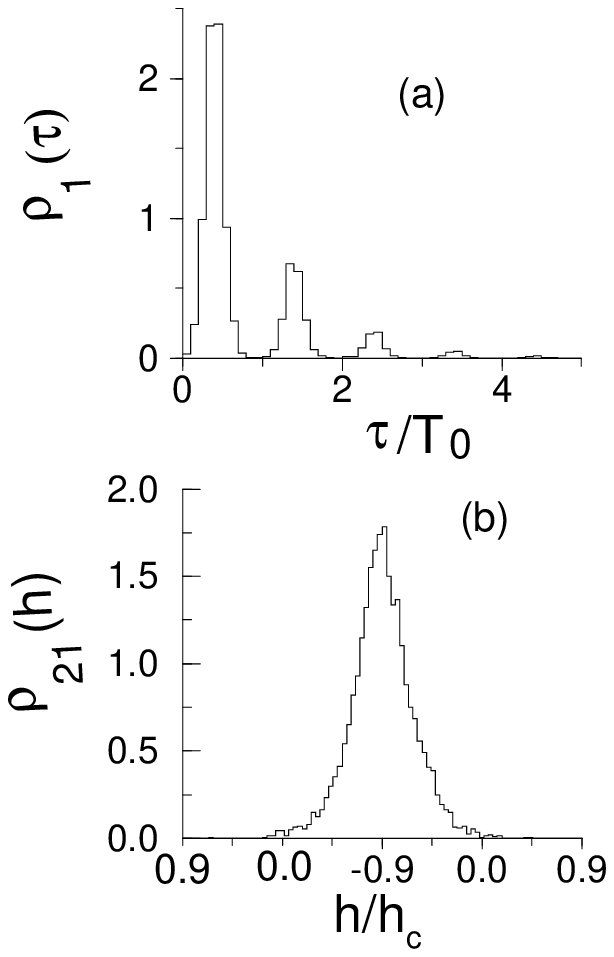}}
\end{figure}
\vfill
\newpage
\begin{figure}
\protect\centerline{\epsfxsize=6in \epsfbox{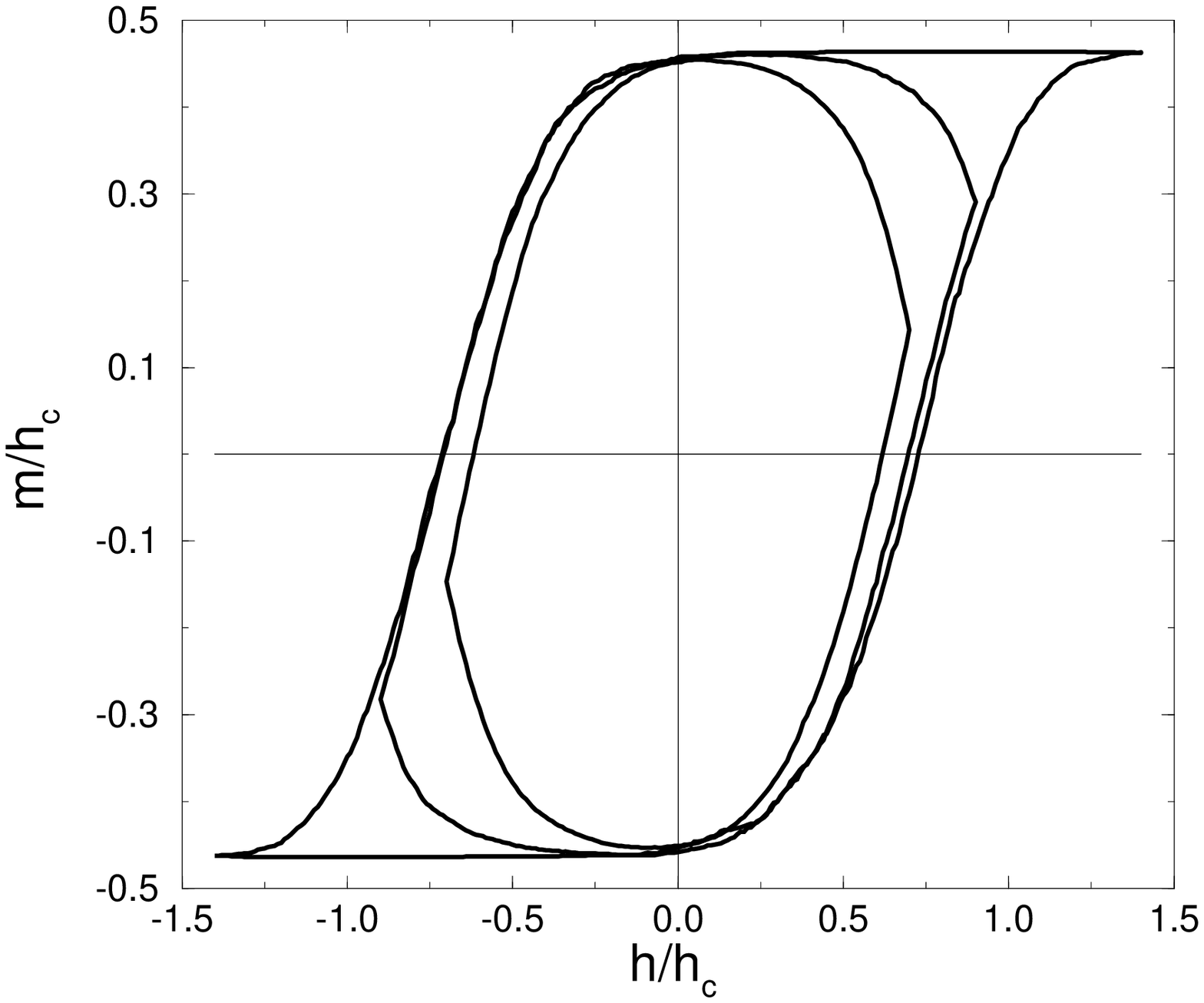}}
\end{figure}
\newpage
\begin{figure}
\protect\centerline{\epsfxsize=6in \epsfbox{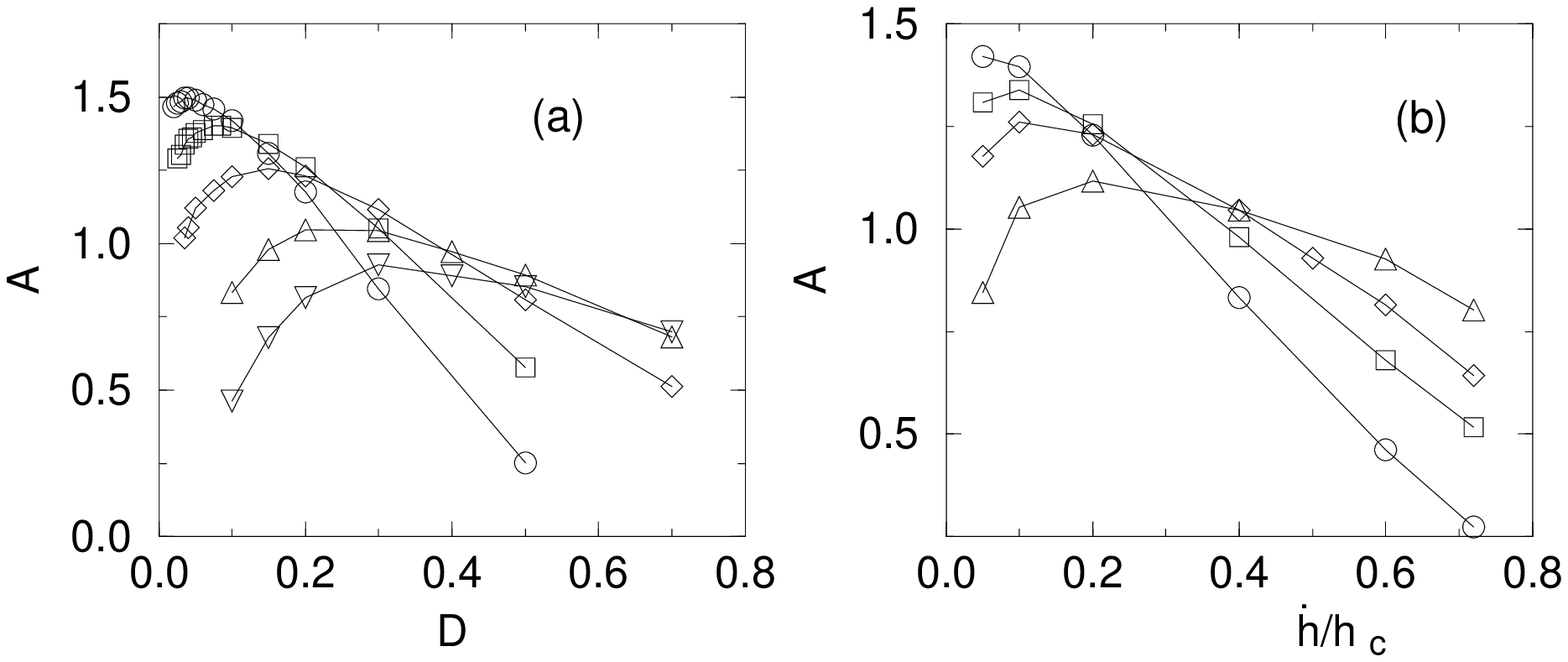}}
\end{figure}
\newpage
\begin{figure}
\protect\centerline{\epsfxsize=6in \epsfbox{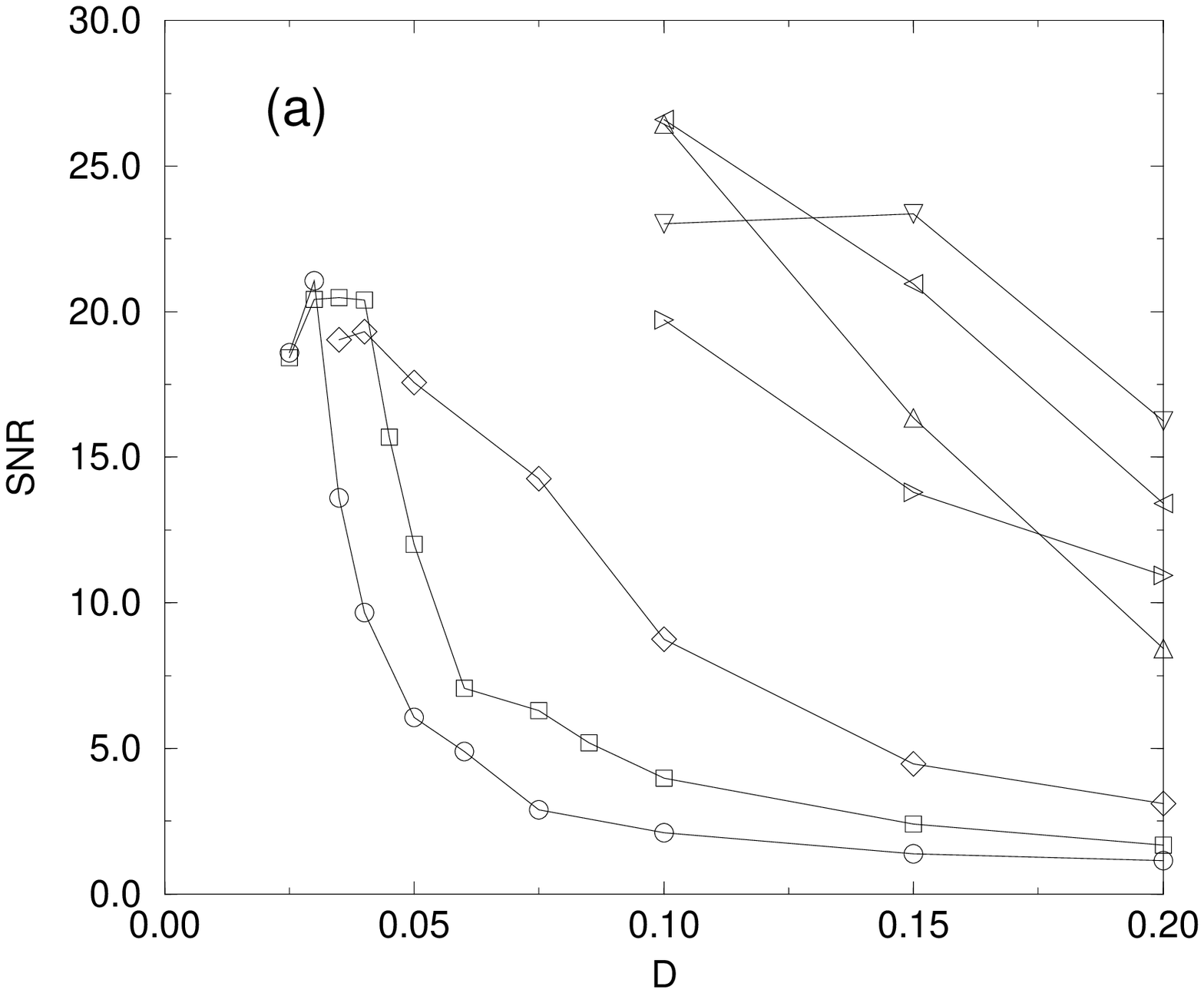}}
\end{figure}
\newpage
\begin{figure}
\protect\centerline{\epsfxsize=6in \epsfbox{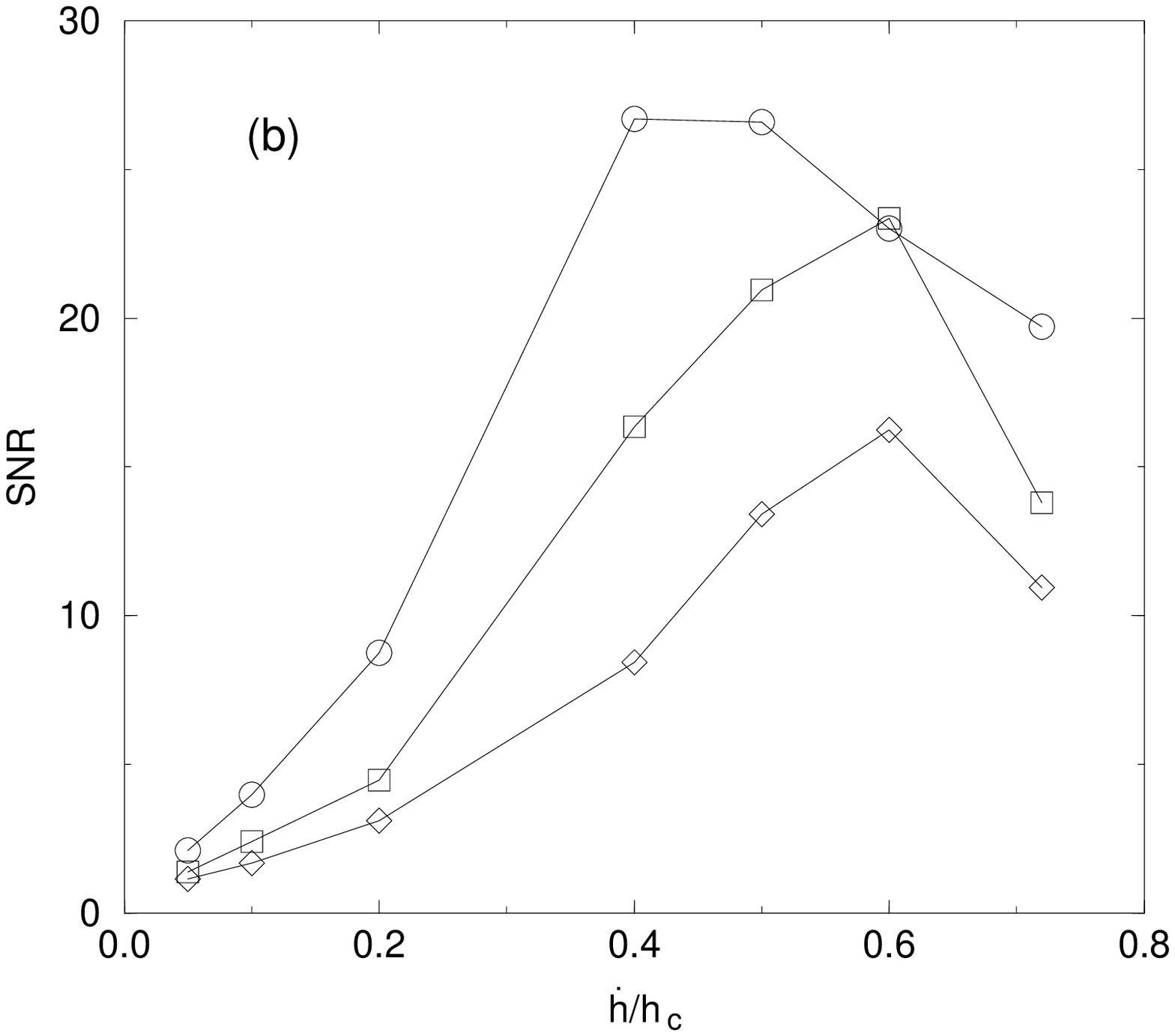}}
\end{figure}
\begin{figure}
\protect\centerline{\epsfxsize=6in \epsfbox{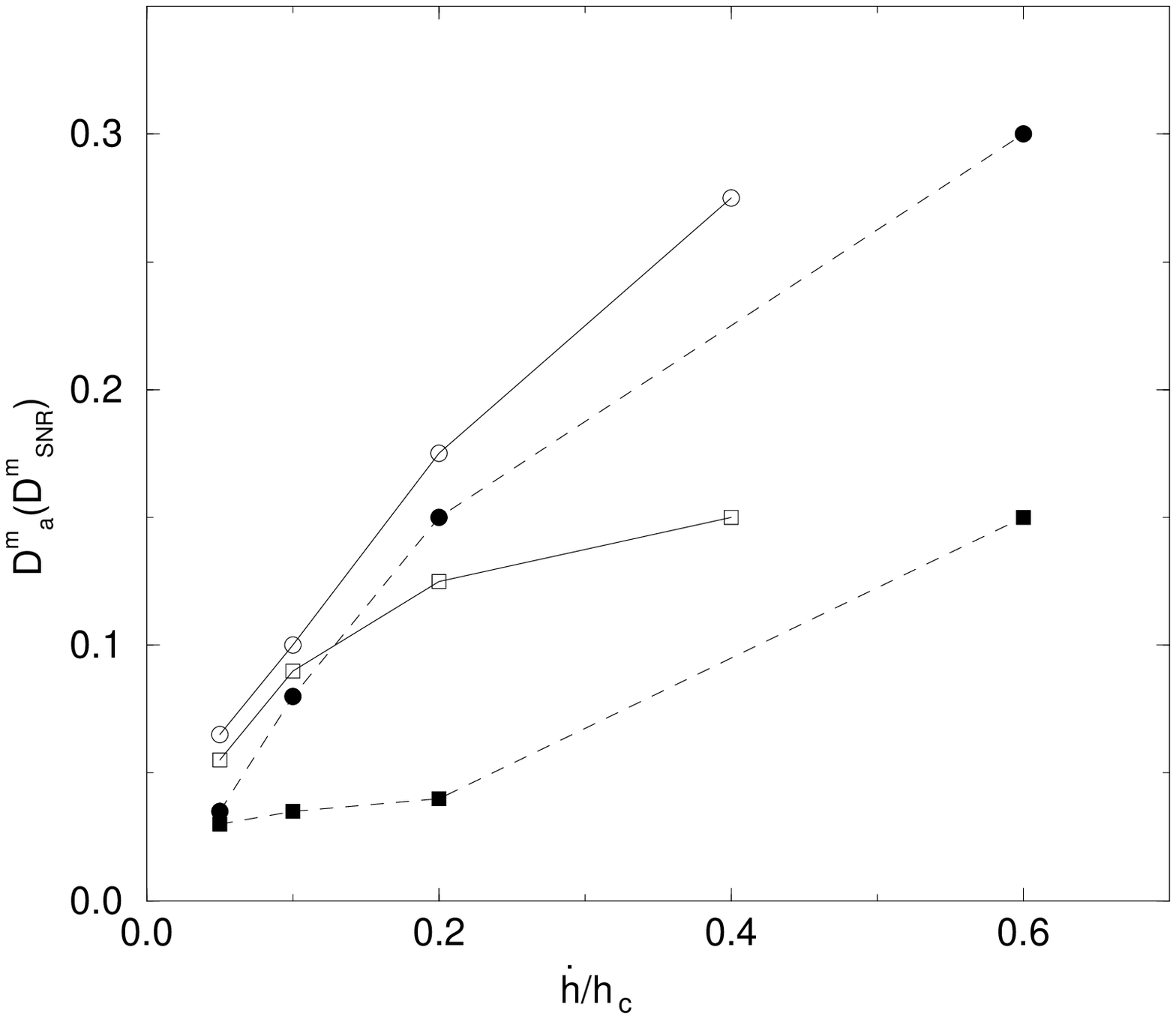}}
\end{figure}

\newpage
\begin{figure}
\protect\centerline{\epsfxsize=6in \epsfbox{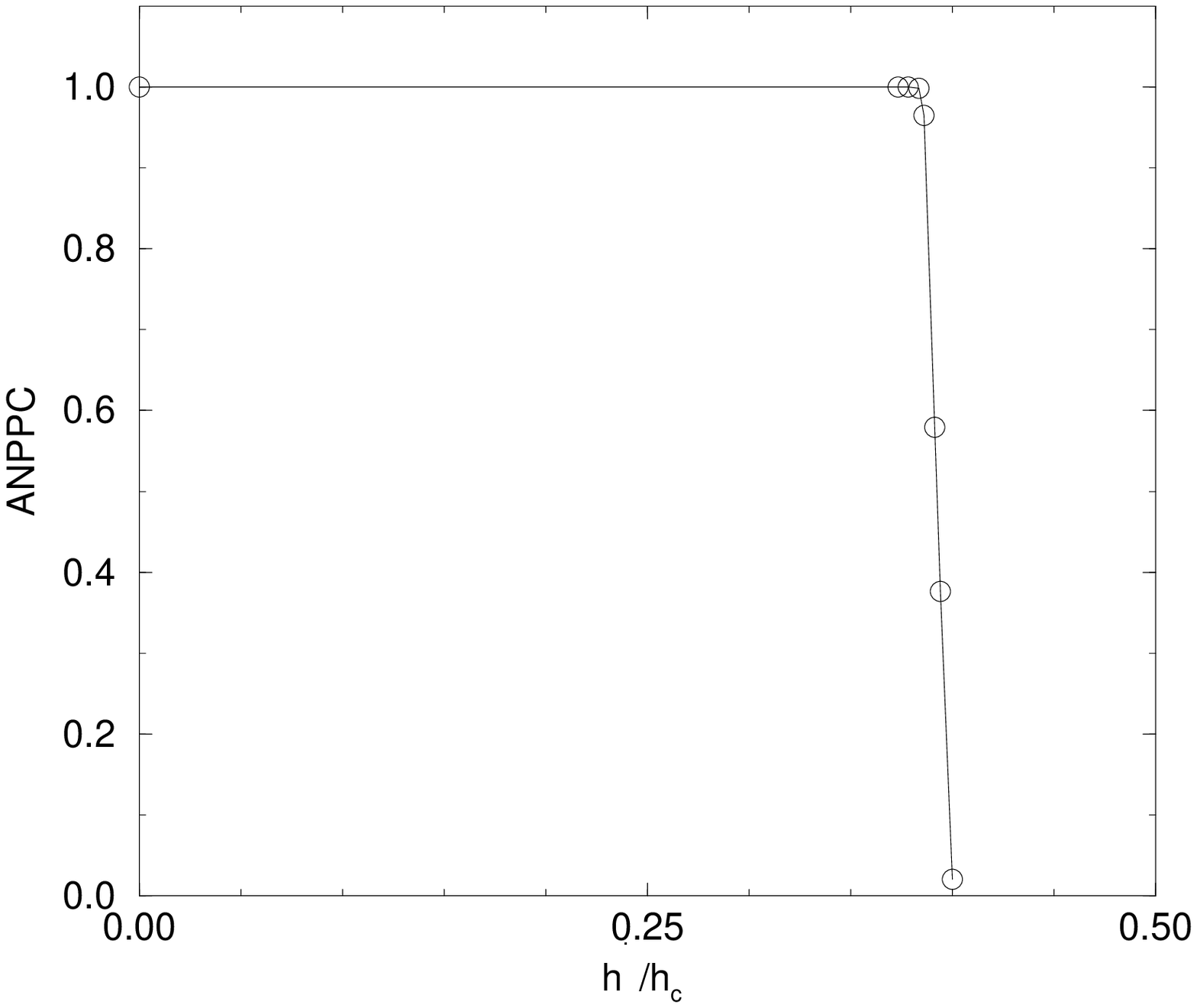}}
\end{figure}
\vfill
\newpage
\begin{figure}
\protect\centerline{\epsfxsize=6in \epsfbox{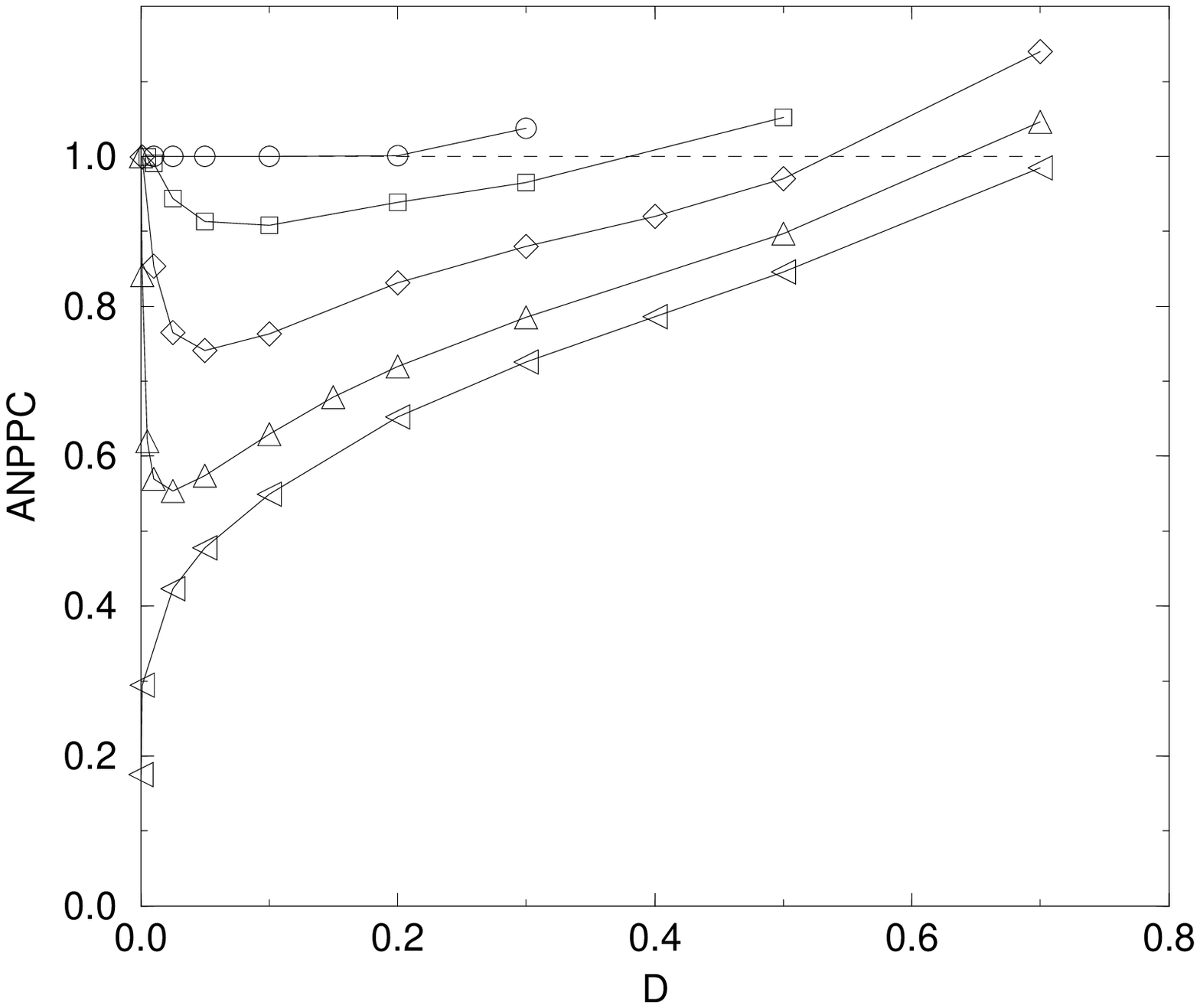}}
\end{figure}
\vfill
\newpage
\begin{figure}
\protect\centerline{\epsfxsize=6in \epsfbox{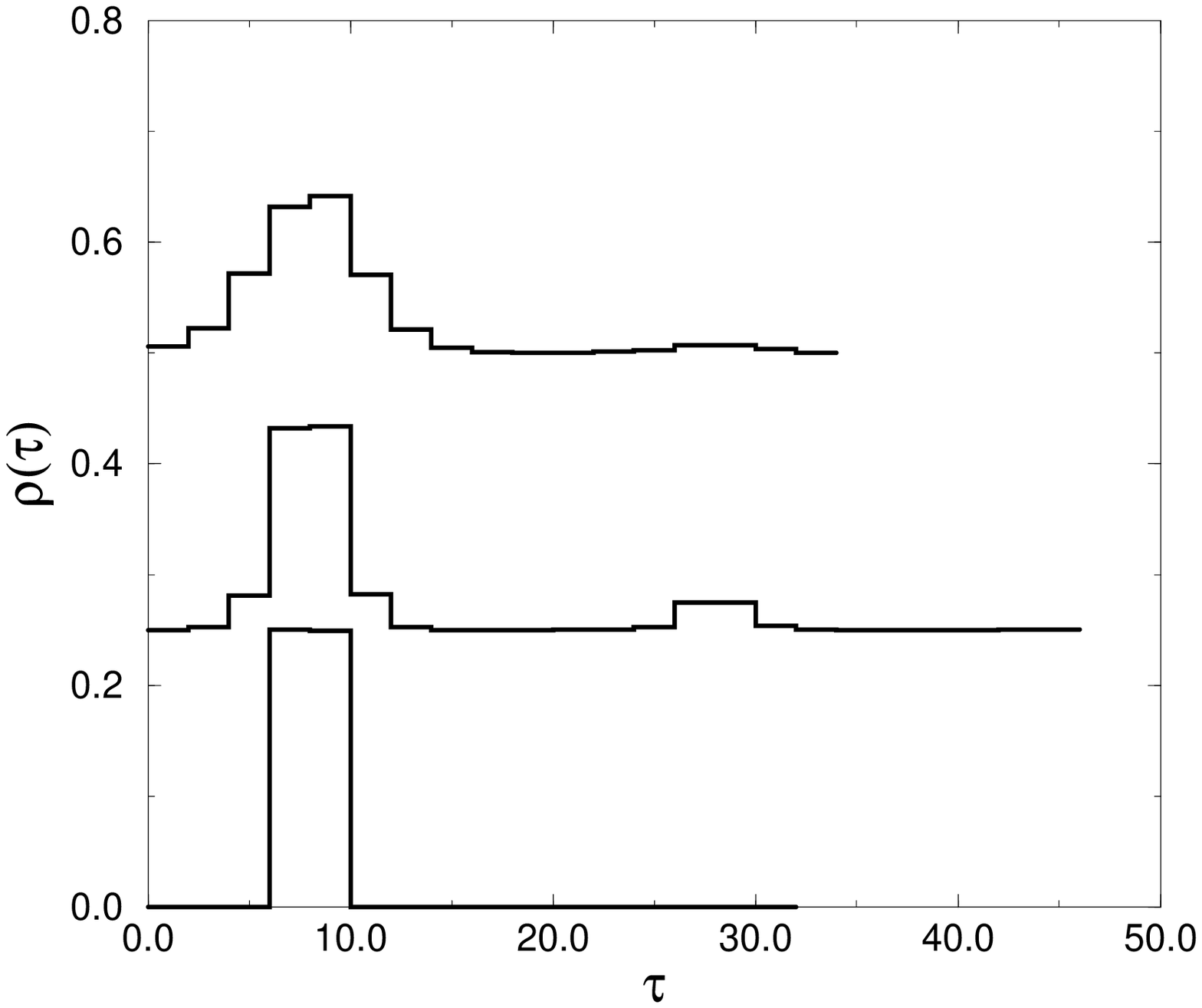}}
\end{figure}
\vfill
\newpage
\begin{figure}
\protect\centerline{\epsfxsize=6in \epsfbox{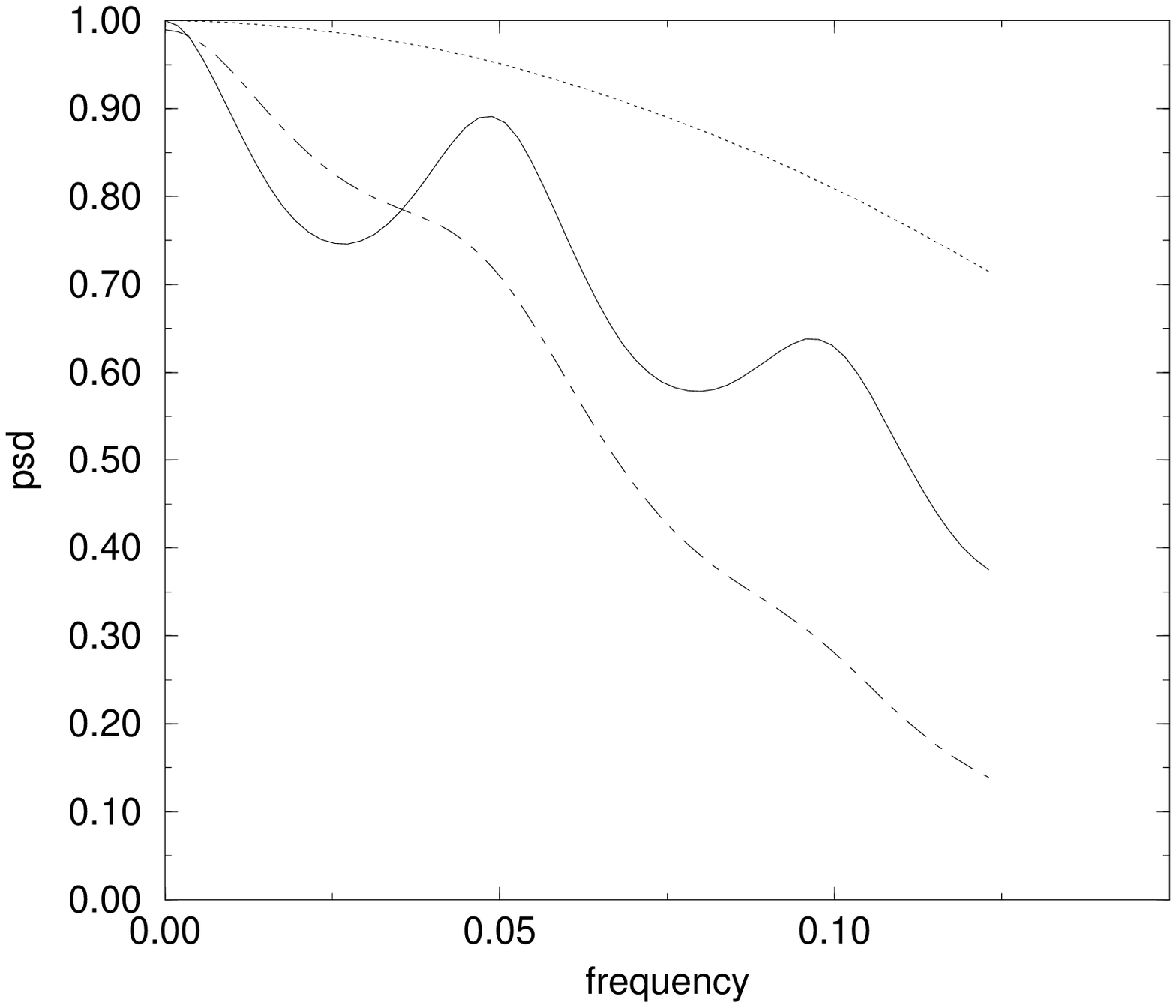}}
\end{figure}
\vfill

\newpage
\begin{figure}
\protect\centerline{\epsfxsize=6in \epsfbox{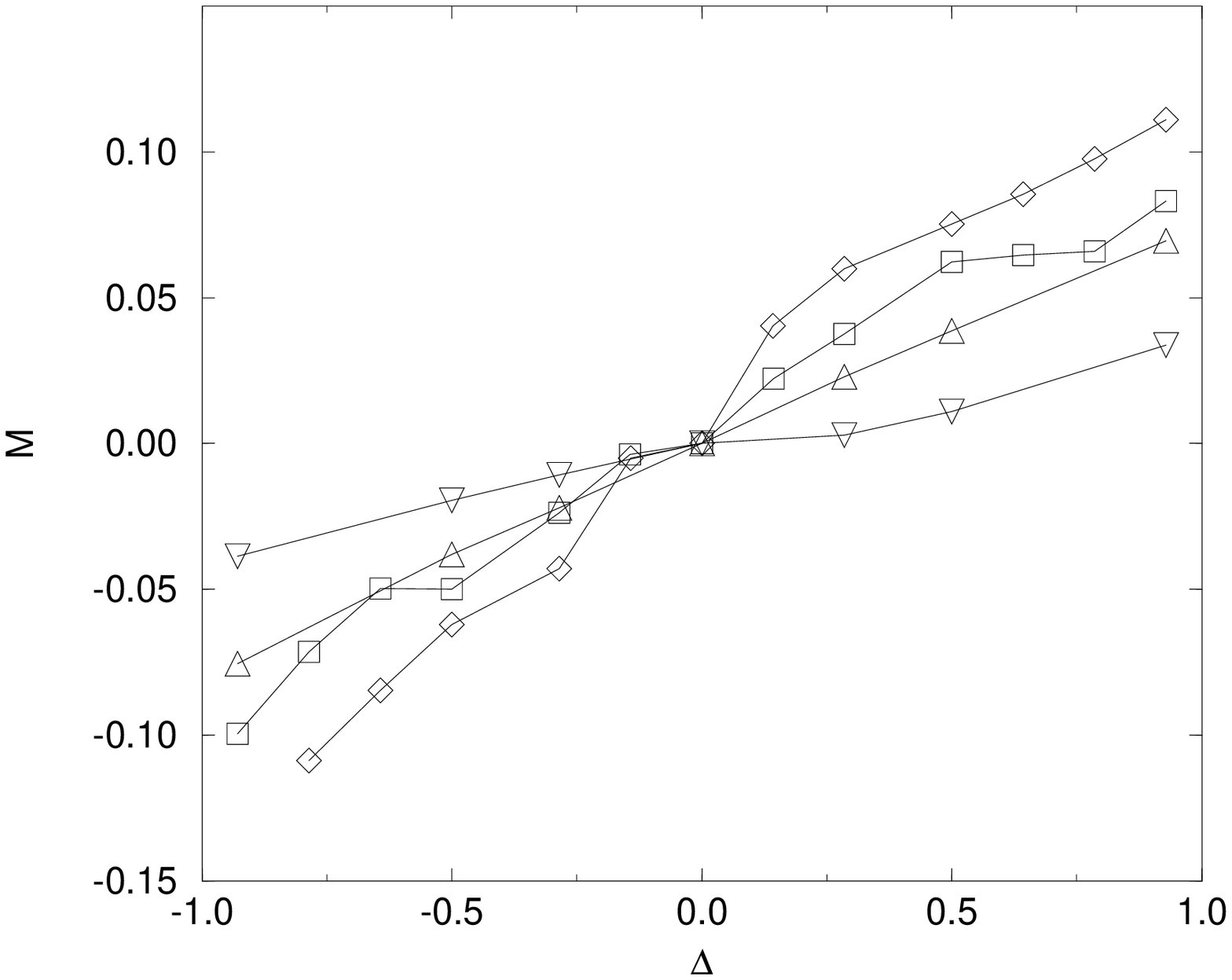}}
\protect\bigskip
\end{figure}
\newpage
\begin{figure}
\protect\centerline{\epsfxsize=6in \epsfbox{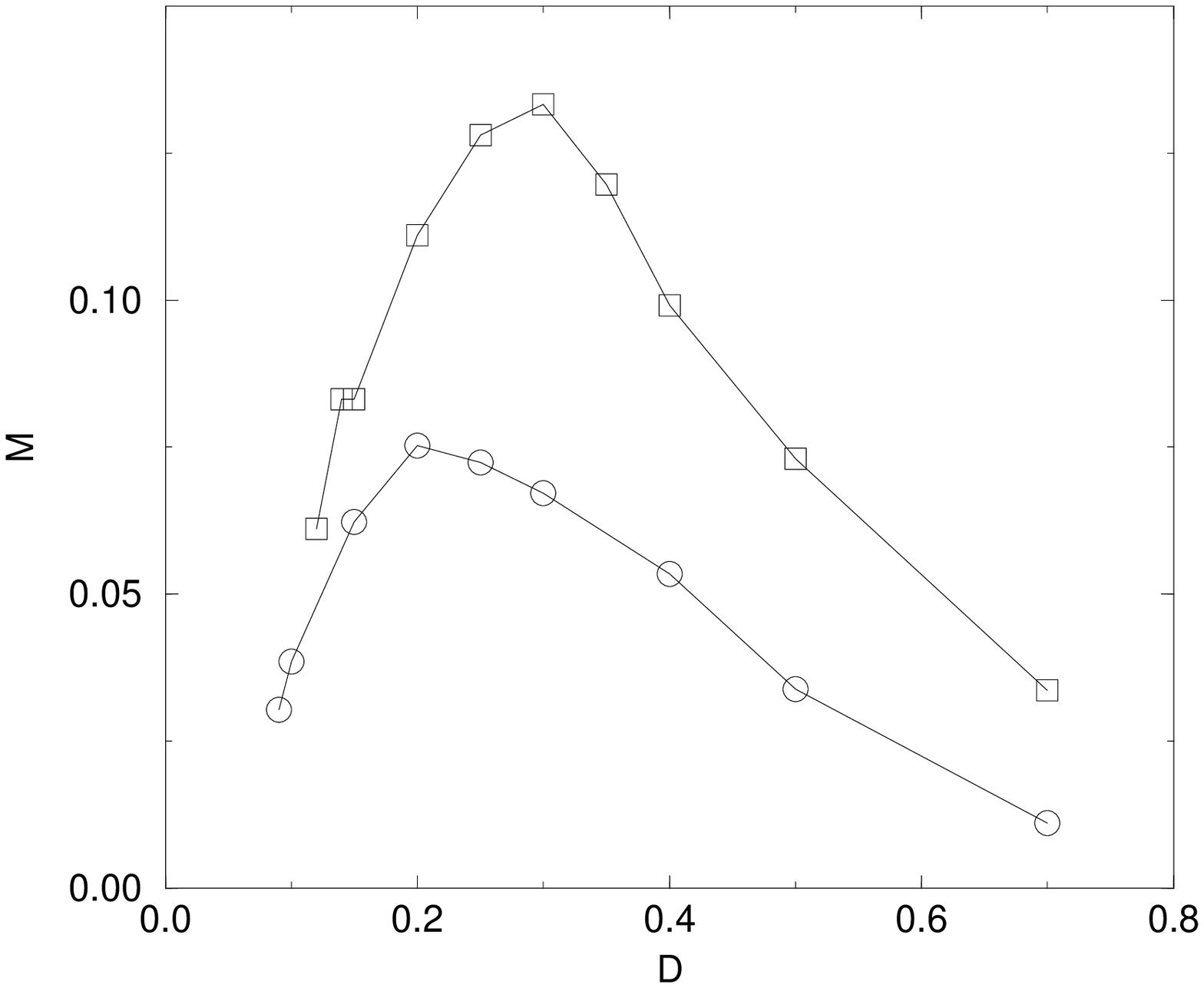}}
\protect\bigskip
\end{figure}
\vfill
\newpage
\begin{figure}
\protect\centerline{\epsfxsize=6in \epsfbox{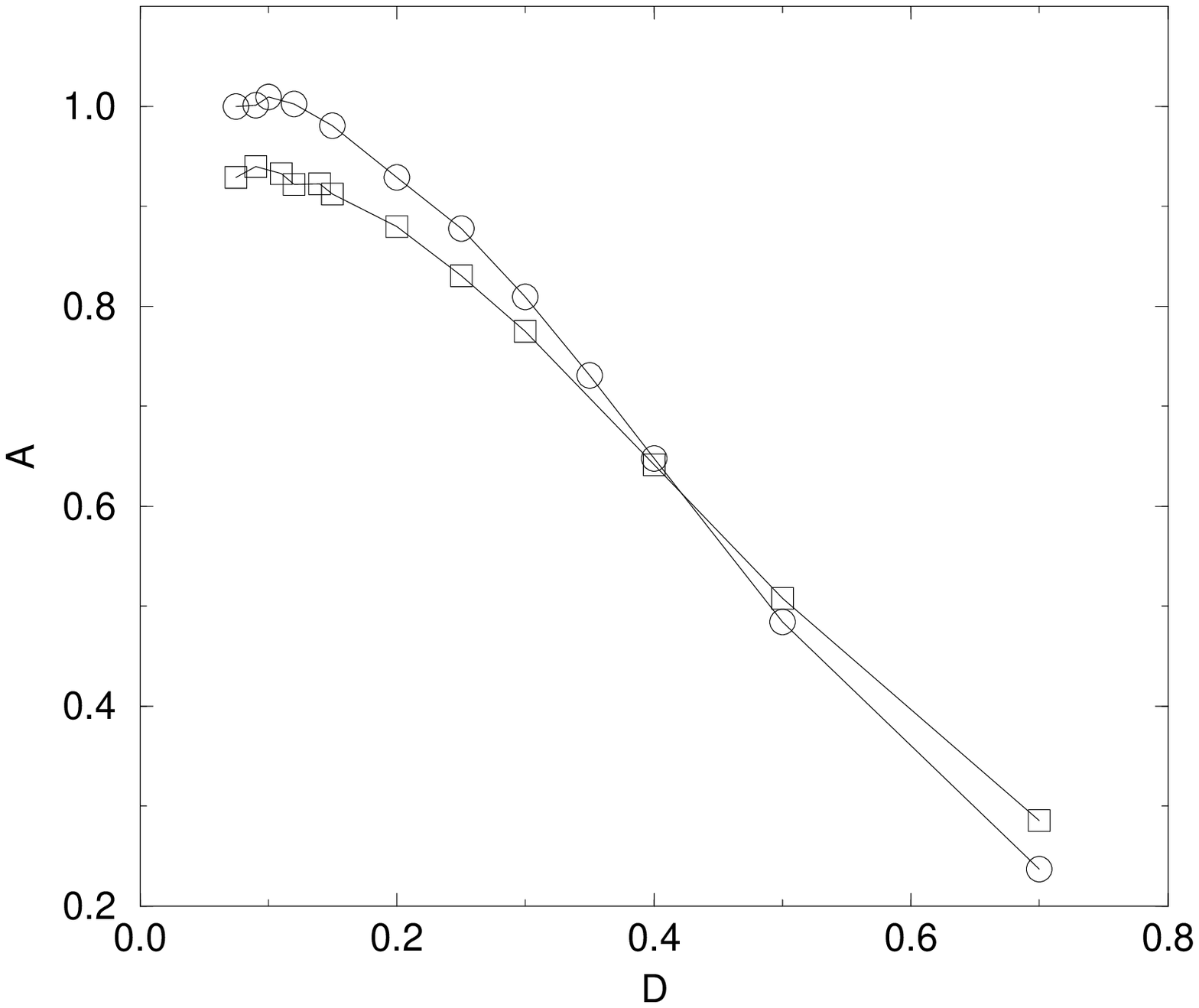}}
\protect\bigskip
\end{figure}
\vfill
\end{document}